# Demographic Inference from Social Media Data with Multimodal Foundation Models: Strategies, Evaluation, and Benchmarking

Hao Yang[1], Angela Yao[*1], Eric Chang[2], and Hexiang Wang[3]

[1]Department of Geography, University of Georgia, Athens, GA, USA

[2]Computer Science Department and Math Department, Georgia Institute of Technology, Atlanta, GA, USA

[3]College of Environment and Design, University of Georgia, Athens, GA, USA

[*]Corresponding Author: Angela Yao      Email: xyao@uga.edu

## 1   Introduction

Contemporary social media platforms enable users to generate and share vast amounts of behavioral data and opinions at an unprecedented scale and granularity. These social media data have been increasingly used in research across various domains, including public opinion analysis(Dong & Lian, 2021; Tavoschi et al., 2020), political communication(Liu et al., 2021), disaster response(Houston et al., 2015), health surveillance(Skaik & Inkpen, 2021; Yang et al., 2023), misinformation detection(Shu et al., 2017), and event monitoring(Mredula et al., 2022). However, using these data to understand social phenomena at the population level remains challenging due to their inherent non-representativeness and sampling bias. Social media users rarely disclose explicit demographic attributes such as age, gender, or race. Without this contextual information, analyses of online attitudes, opinions, and behaviors can be biased toward dominant or more vocal groups, obscuring systematic differences across social strata.

Demographic inference serves as a bridge between digital traces and the underlying populations they represent. By estimating demographic characteristics from users' linguistic patterns, profile information, and behavioral signals, researchers can assess whose voices are being captured in online conversations and how opinions differ across social groups, enabling a more nuanced understanding of social dynamics.

Current methods for demographic inference rely on a variety of information sources, including names(Jaech & Ostendorf, 2015; Mislove et al., 2011), text-based features (Chekili & Hernandez, 2024; Nguyen et al., 2014), social connections and interaction patterns(Al Zamal et al., 2012; Chen et al., 2015), and profile images(Wang et al., 2019), to infer demographic traits such as age, gender, and race. However, these approaches have notable limitations. Most existing models still depend on a single modality of source data, such as text, image, or network, restricting their ability to capture the complex and multimodal nature of online identity expression. In addition, current methods typically infer only one or two demographic attributes (e.g., age or gender) rather than predicting multiple demographic dimensions simultaneously. Finally, many machine learning–based models require large amounts of labeled data, which are costly to obtain, ultimately constraining their generalizability, fairness, and robustness across diverse populations and social contexts.

The rapid development of multimodal foundation models, such as GPT, Gemini, and Claude, has opened new opportunities for advancing demographic inference(Lyu et al., 2025). Trained on massive corpora of words, images, audio, or codes, these models reflect diverse social, cultural, geographical, and linguistic patterns. During training, they implicitly learn statistical associations between linguistic features, such as word choice, syntax, and topics, and demographic or sociolinguistic characteristics. Beyond textual content, user-level information such as usernames, profile descriptions, emojis, and images also carry rich demographic cues. Multimodal foundation models can jointly capture and interpret these textual, symbolic, and visual signals, thereby enhancing the accuracy, robustness, and fairness of demographic inference.

Despite their great potential, the application of foundation models to demographic inference from social media remains largely unexplored. This paper is the first to systematically evaluate the use of multimodal foundation models (e.g., GPT) for inferring age, gender, and race from social media data. We comprehensively assess their performance and accuracy across different input conditions, providing new insights and paving the way for the broader application of multimodal foundation models in demographic inference and computational social science research.

## 2 Theoretical Framework

Demographic inference from social media profiles builds on the understanding that individuals often express aspects of their identity, intentionally or unintentionally, through linguistic, visual, and behavioral cues. As shown in Figure 1, multimodal foundation models (e.g., GPT-5, Gemini, Claude-Opus) enable joint reasoning across heterogeneous information sources, such as text, imagery, and metadata, to infer latent demographic attributes including age, gender, and race. These models are pretrained on vast corpora spanning multiple modalities, learning to represent complex semantic relationships between words, visual concepts, and contextual features. When applied to social media data, these models encode user-specific information, such as usernames, display names, textual posts, and profile images, into a shared latent space where demographic indicators can emerge through correlation patterns. This section outlines the theoretical foundations of how such models capture demographic signals from each modality and integrate them within a unified reasoning framework.

### 2.1 Linguistic and Naming Cues

On social media platforms, usernames and display names, though not always real names, carry rich sociolinguistic signals linked to demographic characteristics.

**Age.** Naming conventions and styles evolve across generations, reflecting cultural trends, media influence, and linguistic innovation(Johfre, 2020). As such, usernames and display names can serve as temporal markers of generational identity. For example, names common in earlier decades (e.g., Deborah, Linda, Michael, Robert) are often associated with older cohorts, whereas more recent generations favor names that emerged in the late 1990s and 2000s (e.g., Aiden, Chloe, Jayden, Isabella). Younger users also tend to adopt creative or playful usernames incorporating abbreviations, numerals, or emojis (e.g., xX_LunaXx, NoCapEthan), signaling digital nativeness and participation in online subcultures. Older users, by contrast, more often employ full given names or professional identifiers (e.g., Robert_Smith, LindaJohnson1965). Foundation models can internalize such generational shifts through co-occurrence patterns between name forms, orthographic conventions, and cultural contexts, thereby aiding in age inference.

**Gender.** Names also carry gendered conventions deeply embedded in linguistic and cultural norms(Espinosa & Xiao, 2018; Pilcher, 2017; Santamaría & Mihaljević, 2018). Large-scale pretraining allows multimodal models to learn probabilistic associations between name morphology and gender identity. For instance, suffixes such as "-a," "-ia," or "-ine" often appear in feminine names, while "-o," "-us," or "-an" are more common in masculine ones. Usernames may further include explicit gender markers—pronouns ("he/him," "she/her"), role identifiers ("Mr.," "Queen"), or gendered nicknames (e.g., QueenBae, Mr.Smith). Beyond naming, stylistic patterns such as emoji use, capitalization, or tone also contribute gendered cues that models can capture and contextualize.

**Race and Ethnicity.** Names encode cultural, linguistic, and geographic signals that can correlate with race and ethnicity(Kozlowski et al., 2022; Lockhart et al., 2023). First names and surnames often originate from specific linguistic or cultural traditions—e.g., Nguyen, Tran, Le (Vietnamese); Hernández, Rodríguez, Lopez (Hispanic/Latino); Wang, Zhang, Li (Chinese); Oluwaseun, Adeyemi (West African); O'Connor, Murphy (Irish). Models trained on diverse global corpora can recognize co-occurrence patterns between such name components and cultural contexts, supporting probabilistic inference of ethnic or cultural background.

## 2.2 Textual Behavior and Stylistic Expression

The linguistic content and style of social media posts convey behavioral and cultural cues tied to demographics through vocabulary, syntax, topic preferences, and paralinguistic markers.

**Age.** Younger users often employ evolving internet slang, abbreviations, memes, and emojis, reflecting digital-native communication norms. Older users may prefer more formal syntax, punctuation, and longer sentences(Pfeil et al., 2009; Rickford & Price, 2013; White et al., 2018). Topic preferences, such as pop culture versus family or professional matters, also vary across age groups. Foundation models trained on large-scale social text corpora capture these generational distinctions by encoding co-occurrence patterns between linguistic features, discourse style, and thematic focus.

**Gender.** Gendered communication patterns emerge in both content and tone(Bamman et al., 2014). On social media, topic preferences often differ across gender lines: female users may post more about lifestyle, emotions, fashion, or relationships, while male users tend to focus on sports, politics, or technology. Beyond topic choice, stylistic differences, such as expressive punctuation, emotive language, and emoji usage, also serve as cues. Models pretrained on diverse text data can implicitly learn these stylistic and topical distributions.

**Race and Ethnicity**. Linguistic expression also reflects culturally rooted communication styles. For instance, African American English (AAE) incorporates characteristic lexical and grammatical forms ("finna," "nah," "ain't"), while code-switching or hybrid forms (e.g., Spanglish) appear in bilingual communities. Asian users may include transliterated words, honorifics, or culturally specific references (e.g., "oppa," "namaste," "kawaii"). Topic preferences may also differ across cultural groups, influenced by community concerns and identity expression, such as discussions of social justice, immigration, cultural festivals, or diasporic belonging(Bindra & DeCuir-Gunby, 2020; Senft & Noble, 2013). Foundation models equipped with contextual embeddings can capture these linguistic variations, associating them with cultural and ethnic backgrounds.

## 2.3 Visual Cues

Profile images, whether selfies, portraits, or symbolic pictures, encode substantial visual information about users' demographic characteristics such as age, gender and race(Cao et al., 2018; Han et al., 2015; Jung et al., 2018).

**Age.** Visual self-presentation often reflects generational preferences. In personal photos, cues such as facial features, hairstyle, clothing, and gestures can signal age. Even when users select non-personal images, their choices convey age-related tendencies: younger users may prefer icons from popular culture, anime, or gaming, while older users might use family photos, landscapes, or traditional imagery. Foundation models can learn to associate such patterns with generational cohorts to infer approximate age groups.

**Gender.** Profile images often exhibit gendered self-representation. Selfies and portraits may reveal facial morphology, hairstyle, clothing, and makeup styles, while symbolic imagery (e.g., flowers, pets, lifestyle aesthetics for females; cars, sports icons, or abstract designs for males) can also reflect gendered preferences, though these associations are not deterministic. Multimodal models integrate both direct visual and contextual cues to support gender inference.

**Race and Ethnicity.** Visual representations can signal racial or ethnic identity directly, through facial features, skin tone, or attire, or indirectly, through symbolic or cultural imagery such as religious icons, national flags, or heritage motifs. Vision transformers within multimodal foundation models can extract such visual patterns and associate them with learned representations of racial and cultural diversity.

Theoretically, foundation models capture these demographic signals probabilistically rather than deterministically, identifying linguistic, phonetic, and visual patterns that correlate with certain age, gender, and ethnic groups while maintaining awareness of uncertainty. The strength of multimodal foundation models lies in their ability to integrate cross-modal cues, linking linguistic features with visual representations and user metadata. For instance, when a username suggests one gender but the profile image indicates another, the model can weigh evidence probabilistically to reach a balanced inference. Attention mechanisms allow selective focus on salient features, while joint embedding spaces enable knowledge transfer across modalities. This integrative reasoning mirrors human perception, understanding identity through both what people say and how they present themselves.

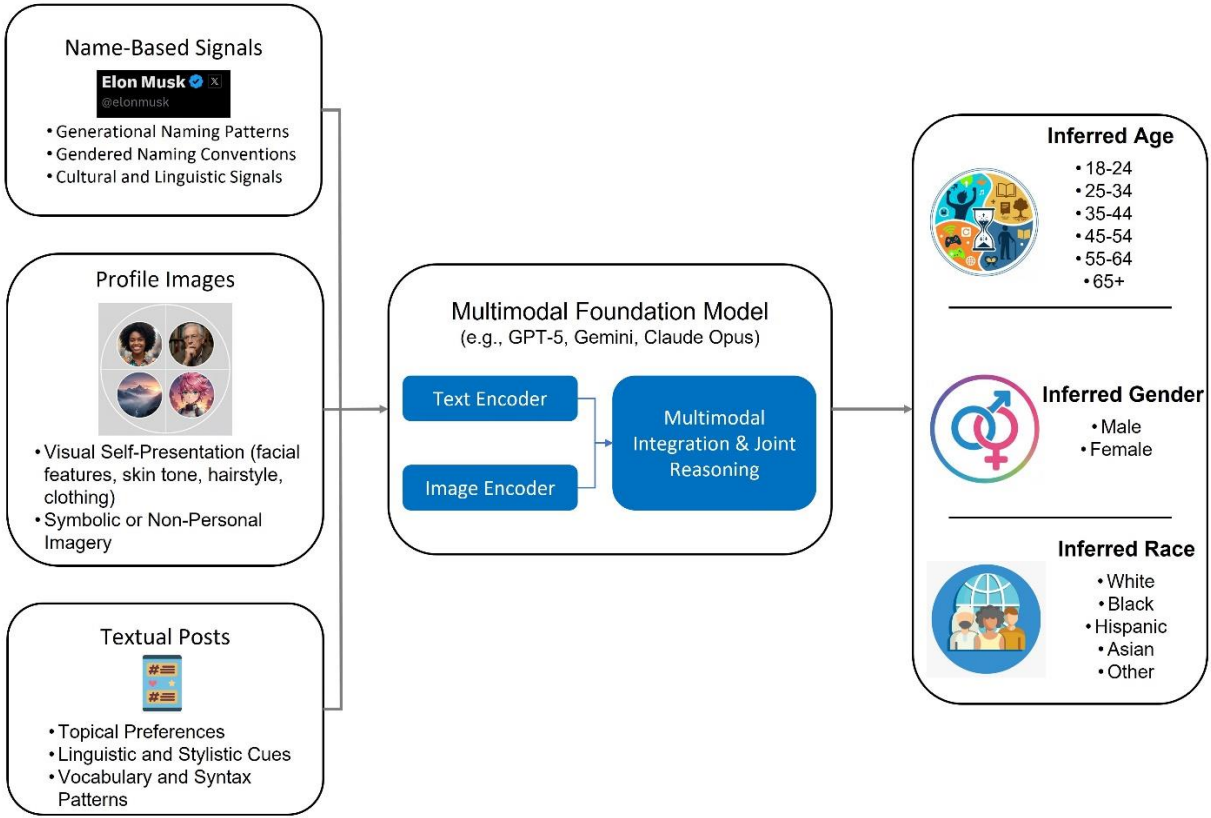

Figure 1. Demographic signals embedded in multimodal social media inputs

## 3 Methodology

This study leverages the capability of GPT-5, a state-of-the-art multimodal foundation model, to infer user demographics, including age, gender and race/ethnicity, from social media information from X (formerly Twitter). We developed a workflow to collect a stratified random sample of 263 X user profiles. From this dataset, a subset of 120 user profiles was manually labeled with demographic information, serving as ground-truth for accuracy evaluation of the inference.

The study developed multiple strategies for demographic inference with the use of GPT-5 API and compared their effectiveness. Benchmark evaluation results are presented and several strategic recommendations are summarized. The findings offer insights that can guide future applications of foundation models for demographic inference.

### 3.1 Data

This study focuses on English-language social media data from X, restricted to users based in the United States. A total of 263 user profiles were collected. All information used in this study is publicly available and consistent with X's Terms of Service. The user sample was identified and selected through the following stepwise process:

1. **Initial Candidate Pool**:
   We first utilized an archived dataset of X/Twitter users as a source of candidate accounts and randomly selected users to form the initial pool.

2. **Birthday-based User Search**:
   We then used the native X search interface and keyword filters (e.g., "it's my birthday," "turning 25," "happy 30th to me") to identify users who publicly posted about their own birthdays. These posts provided verifiable temporal anchors for inferring users' ages or birth years.
3. **Inclusion Criteria**:
   To ensure credible demographic ground truth, users were included only if they met all of the following:
   - Age: Clearly stated or verifiable from self-declared birthday posts (e.g., "turning 27 today").
   - Gender: Evident from self-identifying pronouns or selfie presentation.
   - Race/Ethnicity: Inferable from profile images (selfies), tweet posted and broader cultural or linguistic context.
   - Language: The majority of profile content and tweets were in English.
   - Geography: Indications (location field, hashtags, content) suggested residence in the United States.
   
   Users with ambiguous or inconsistent demographic indicators were excluded and replaced with new candidates.
4. **Data Fields Collected**:
   For each selected user, we then extracted Display name, Handle (username), Short bio or profile description, Profile image, and three most recent tweets (excluding retweets or replies). To fairly evaluate the foundation models' capability in demographic inference, the collected tweets did not include birthday-related content or embedded images—only textual content and emojis were retained.

To establish ground truth demographic labels, a four-member trained annotation team with social media research experience conducted manual labeling. Annotators completed a structured training session with detailed coding guidelines defining each demographic category and examples of valid evidence.

Each user profile was independently reviewed by two annotators, and disagreements were resolved through adjudication by a third annotator to reach consensus.

- **Age Labeling**: Determined from explicit self-reports (e.g., "turning 25 today") or inferred from time-stamped birthday posts. Users with uncertain or missing age information were excluded.
- **Gender Labeling**: Inferred from pronouns in bios (e.g., "she/her," "he/him"), first-name conventions, and self-presentation in selfies. Ambiguous or gender-neutral profiles were labeled as unknown and omitted from the final dataset.
- **Race/Ethnicity Labeling**: Determined primarily from visible phenotype in publicly posted selfies, corroborated by textual self-identification (e.g., "Black engineer," "Latina mom") or cultural references. When cues were inconsistent or unclear, the profile was marked unknown and excluded to maintain labeling reliability.

Then, after the selection and annotation process, 263 user profiles were retained. This labeled dataset was designed to capture diversity across age, gender, and race/ethnicity, ensuring that multiple demographic groups were well represented. As shown in Figure 2, the dataset exhibits a balanced composition by gender, with male and female users contributing roughly equal proportions. The age distribution is moderately skewed toward younger adults, particularly those aged 18–34, reflecting the demographic structure of active X (Twitter) users, while older age groups are also included to maintain coverage across the life span. The racial and ethnic distribution includes users identified as White, Black, Hispanic, Asian, and Other, with White and Black users forming the two largest groups, followed by Hispanic and Asian users, indicating reasonable racial diversity.

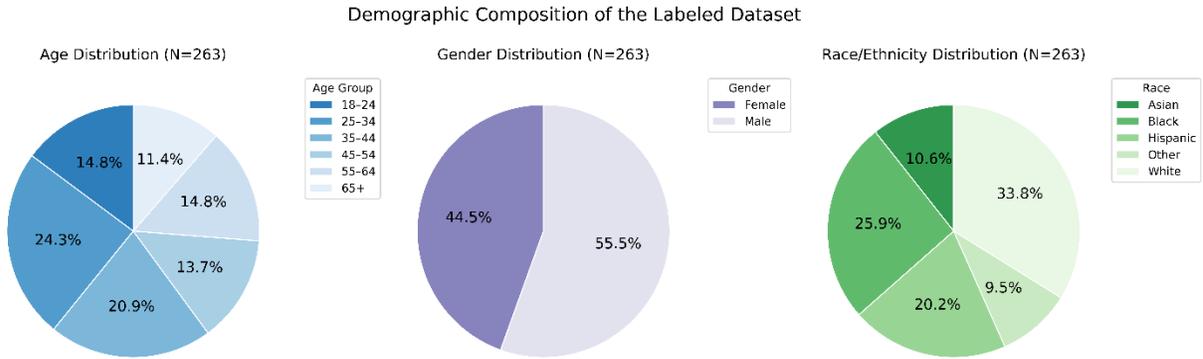

Figure 2. Distribution of users by age, gender, and race/ethnicity in the labeled dataset

As shown in Table 1, the cross-tabulation of age, gender, and race provides a detailed view of the dataset's demographic composition. Each intersectional bin, for example, young Black males or older Asian females, is represented, ensuring that no demographic combination is missing from the labeled data. Although some subgroups (e.g., older Asian or Hispanic users) are relatively small, the dataset overall maintains comprehensive demographic coverage. This balanced yet diverse structure supports fair evaluation of the inference model across both individual and intersectional demographic dimensions.

Table 1. Distribution of Labeled Users by Age Group, Gender, and Race/Ethnicity (N = 263)

| Age Group | Gender | Asian | Black | Hispanic | Other | White | Row Total |
|---|---|---|---|---|---|---|---|
| 18–24 | Female | 2 | 4 | 6 | 3 | 2 | **17** |
| | Male | 2 | 7 | 5 | 2 | 6 | **22** |
| 25–34 | Female | 3 | 7 | 9 | 2 | 11 | **32** |
| | Male | 2 | 16 | 7 | 2 | 5 | **32** |
| 35–44 | Female | 2 | 5 | 5 | 2 | 8 | **22** |
| | Male | 3 | 12 | 4 | 2 | 12 | **33** |
| 45–54 | Female | 2 | 2 | 3 | 3 | 6 | **16** |
| | Male | 3 | 4 | 3 | 2 | 8 | **20** |
| 55–64 | Female | 2 | 3 | 2 | 2 | 7 | **16** |
| | Male | 3 | 3 | 4 | 2 | 11 | **23** |

| | Female | 2 | 2 | 2 | 2 | 6 | **14** |
|---|---|---|---|---|---|---|---|
| **65+** | Male | 2 | 3 | 3 | 1 | 7 | **16** |
| **Column Total** | | 28 | 68 | **53** | **25** | 89 | **263** |

During data selection, annotation, and storage, only publicly accessible information was used, and no attempts were made to contact or interact with users. The dataset was stored securely in accordance with institutional research ethics protocols. This study is classified as minimal-risk human subjects research, as demographic attributes were inferred solely for research benchmarking purposes, not for individual profiling, and no personal opinions or sensitive content were analyzed.

## 3.2 Method

We applied multiple strategies to apply GPT-5 for user demographic inference with consideration of the following factors.

### 3.2.1 Progressive Information Conditions

To examine how varying amounts of profile information influence model performance, we employ a progressive-input evaluation approach. The model is presented with increasingly richer amount of user information, moving from minimal textual input (names only) to the full multimodal setting (including profile images).

This design serves two main purposes. First, it allows us to assess the contribution of each information source, names, descriptive text, linguistic behavior, and visual appearance, to overall inference accuracy. Second, it reflects a realistic human reasoning process, where one begins with name-based heuristics and then incorporates more contextual and visual information. As shown in Table 2, each user profile is evaluated under five sequential conditions:

Table 2. Overview of the five sequential input conditions (C1–C5) defining the progressive multimodal evaluation framework

| Condition | Input Provided to GPT-5 | Purpose |
|---|---|---|
| **C1** | Username/Handle + Display Name | Represents the most limited scenario, where the model relies solely on name morphology, linguistic origin, and cultural naming conventions. This establishes a baseline for minimal-context performance. |
| **C2** | C1 + Short Description | Adds short self-descriptive text, which often includes pronouns, occupations, or interests, allowing the model to capture self-identification cues beyond naming patterns. |
| **C3** | C2 + One Tweet | Expands the evidence to include a user's linguistic style, topical preferences, and expressive elements such as emojis or slang, providing additional implicit demographic signals. |
| **C4** | C3 + Two More Tweets | Extends C3 by adding two more tweets, enabling the model to access richer linguistic and topical variation. |
| **C5** | C4 + Profile Image | Completes the multimodal setting by incorporating visual information, allowing GPT-5 to integrate textual and visual cues within a unified reasoning process. |

All other model parameters are held constant across conditions (temperature = 0.0 to eliminates stochasticity, top-p = 1.0 no artificial probability truncation, max_tokens = 512 to keeps response lengths comparable) to ensure deterministic outputs. This design isolates how incremental information improves demographic inference accuracy.

### 3.2.2 Prompting Strategy

GPT-5 was guided using a Chain-of-Thought (CoT) prompting approach. In each run, the model was instructed to reason step by step about demographic clues, including name structure, pronouns, linguistic style, topics of tweets, and visible attributes in the profile image, before producing a structured JSON output specifying:

- age_group: one of [18–24, 25–34, 35–44, 45–54, 55–64, 65+]
- gender: one of [male, female]
- race_ethnicity: one of [White, Black, Hispanic/Latino, Asian, Other]
- confidence_0to1: a value between 0 and 1

If the available evidence was insufficient, the model could set "abstain": true and assign a low confidence score (< 0.5) with the label "unknown." This prompting framework improves interpretability, reduces random guessing, and enables transparent inspection of how GPT-5 weighs linguistic, behavioral, and visual cues when inferring demographic attributes.

Building on this reasoning framework, the five progressive prompting conditions (C1–C5) were designed to evaluate how the inclusion of additional information affects GPT-5's inference accuracy and reasoning behavior. Each condition uses a corresponding prompt template tailored to its input configuration. The templates share a consistent structure, requiring step-by-step reasoning and structured JSON output, but differ in their instructions regarding which information sources are available and relevant.

As illustrated in Figure 3, the prompt for Condition C1 (names only) elicits GPT-5's reasoning process for each demographic dimension, along with the corresponding confidence scores. Figure 4 presents the adapted prompt for Condition C5, which includes names, description, tweets, and image inputs. In this condition, the prompt explicitly emphasizes cross-modal reasoning and the relative weighting of different cues to enhance inference performance. Since the GPT-5 API currently does not support image inputs, we used GPT-4o via the OpenAI API to perform the multimodal inference for C5, ensuring consistent prompt structure and evaluation settings across all conditions.

By comparing performance across these five conditions, we can quantify the marginal value of additional information and examine how GPT-5's reasoning behavior evolves as the profile context becomes richer. This progressive prompting framework thus offers a controlled and interpretable way to study multimodal reasoning in large foundation models applied to demographic inference.

**Prompt:**
*You are a careful and impartial annotator conducting demographic inference for X (formerly Twitter) users. Your task is to reason step by step about the demographic clues available in each user's username and display name, and then provide a structured output. You are provided with the following information: Username/Handle Display Name Infer the user's age group, gender, and race/ethnicity based on name morphology, cultural or linguistic cues, and, when applicable, emojis or textual symbols that appear in the name. Think carefully before answering: Identify specific evidence for each demographic dimension. Weigh multiple clues if present, and consider the cultural or linguistic context. If evidence is insufficient or ambiguous, choose "unknown" and set "confidence_0to1" below 0.5 rather than guessing. Your structured output should include: "age_group": one of [18–24, 25–34, 35–44, 45–54, 55–64, 65+, unknown] "gender": one of [male, female, unknown] "race_ethnicity": one of [White, Black, Hispanic/Latino, Asian, Other, unknown] "confidence_0to1": number between 0 and 1.*
*First user: username: ▓▓▓▓▓▓▓▓, display name: ▓▓▓▓.*

**GPT-5:**
Here's the structured demographic inference for the user @▓▓▓▓▓▓ | ▓▓▓▓▓ following your specified schema and reasoning protocol:

**Reasoning:**

- **Username ("▓▓▓▓▓▓▓")** provides no clear demographic cue—likely stylized, non-semantic handle.
- **Display name ("▓▓▓▓▓")** is a modern English given name.
    - **Gender:** "▓▓▓▓▓" is overwhelmingly used as a **female** first name.
    - **Age:** The name gained strong popularity in the **late 1990s–2010s**, suggesting a person most likely in the **18–34** range.
    - **Race/Ethnicity:** The name "▓▓▓▓▓" is most common among **White** users in U.S. naming datasets (SSA records), though occasionally appears across groups.

Evidence strength: name-based inference is moderately strong for gender, somewhat moderate for age, and weak-to-moderate for race.

**Structured Output:**

{
 "age_group": "25–34",
 "gender": "female",
 "race_ethnicity": "White",
 "confidence_0to1": 0.75
}

Figure 3. Example of GPT-5 prompt and output under Condition C1 (names only)

**Prompt:**

You are a careful and impartial annotator conducting demographic inference for X (formerly Twitter) users. Your task is to **reason step by step** about the demographic clues available in each user's profile and content, and then provide a structured output.

You are provided with the following information for each user:

- **Username / Handle**
- **Display Name**
- **Short Description (Bio)**
- **Three Tweets**
- **Profile Image**

Based on these inputs, **infer the user's age group, gender, and race/ethnicity** by analyzing:

- Name morphology and linguistic form (e.g., suffixes, conventions)
- Language style, grammar, and vocabulary choices
- Topics, interests, and tone of tweets
- Cultural, regional, or linguistic cues
- Emojis, symbols, or stylistic markers in text
- Visual cues from the profile image related to age, gender, and race/ethnicity

Think carefully before answering:

1. Identify specific evidence supporting each demographic inference.
2. **Weigh multiple clues** when present, considering cultural and linguistic context.
3. If evidence is ambiguous or insufficient, return "unknown" rather than guessing.

Your structured output should follow this format:

{
  "age_group": "one of [18–24, 25–34, 35–44, 45–54, 55–64, 65+, unknown]",
  "gender": "one of [male, female, unknown]",
  "race_ethnicity": "one of [White, Black, Hispanic/Latino, Asian, Other, unknown]",
  "confidence_0to1": "a number between 0 and 1 indicating overall confidence"
}

Figure 4. Input prompt template under Condition C5, incorporating names, profile description, tweets, and image for multimodal inference

### 3.3 Evaluation

Model predictions were compared against human-annotated ground truth labels. For each demographic attribute, we computed accuracy, precision, recall, and F1-score to quantify classification performance. Precision measures the proportion of correctly predicted instances among all predictions of a given class, whereas recall measures the proportion of correctly identified instances among all true cases of that class. The F1-score, the harmonic mean of precision and recall, provides a balanced measure of the model's ability to correctly identify each class. Because both age group and race/ethnicity contain multiple subcategories, metrics were calculated on a per-class basis and then aggregated using macro-averaging, which computes the unweighted mean across all classes, provides a balanced measure of performance.

For age groups, an ordinal variable, we additionally report one-off accuracy, which considers predictions in adjacent categories (e.g., predicting 25–34 instead of 18–24) as nearly correct, and mean absolute error (MAE), computed after mapping each age group to its numeric midpoint. We also report joint accuracy across all three demographic dimensions, where a prediction is counted as correct only if all three labels—age, gender, and race/ethnicity—match the ground truth. Finally, under the progressive-input evaluation, we analyze how model performance changes across the five conditions to quantify the incremental benefit of additional information.

## 4  Results and Discussions

We apply the progressive-input framework to evaluate GPT models' capability in demographic inference using the 263 X user profiles. The analysis begins with the name-only baseline and incrementally incorporates additional profile information, short descriptions, tweets, and profile images, to assess how each input modality contributes to model performance.

### 4.1  Condition 1 (C1): Names Only

In the first condition, the model receives only the username and display name. Table 3 summarizes the metrics for the name-only setting. Age prediction shows limited accuracy (35%), although the one-off accuracy rises to 77%. The mean absolute error (MAE) of approximately 10 years and macro-F1 of 0.32 indicate substantial overlap among adjacent age groups. Gender prediction performs strongly (accuracy = 0.94; macro-F1 = 0.94), reflecting that given names often encode clear binary gender markers. Race/ethnicity inference achieves moderate results (accuracy = 0.69; macro-F1 = 0.59), showing partial separability but persistent confusion between Hispanic and White users and between Asian and "Other" categories. The joint accuracy is only 0.23, highlighting that strong single-attribute performance does not directly translate to overall demographic accuracy.

The confusion matrices (Figure 5) show substantial misclassification among adjacent age cohorts, particularly between 18–24 and 25–34, and within mid-aged groups (55–64). This confirms that names alone provide limited generational information, producing broad-band errors consistent with the 10-year MAE. Gender inference demonstrates robust separability (≈94% accuracy) with minimal cross-gender errors, driven primarily by linguistic morphology and cultural name frequency. Race prediction is moderately successful: White and Black users are relatively well identified, while Asian and Hispanic users experience greater cross-group confusion. This likely reflects the influence of linguistic assimilation and anglicization in online naming practices. Because this study focuses on English-language names, many individuals from non-English-speaking or multicultural backgrounds adopt English-transliterated or culturally neutral names, reducing the distinctiveness of ethnic naming patterns. Consequently, the model struggles to capture subtle ethnocultural cues based solely on name morphology. Overall, the name-only baseline reveals that gender information is highly encoded in user names, race cues are partial, and age cues are weak.

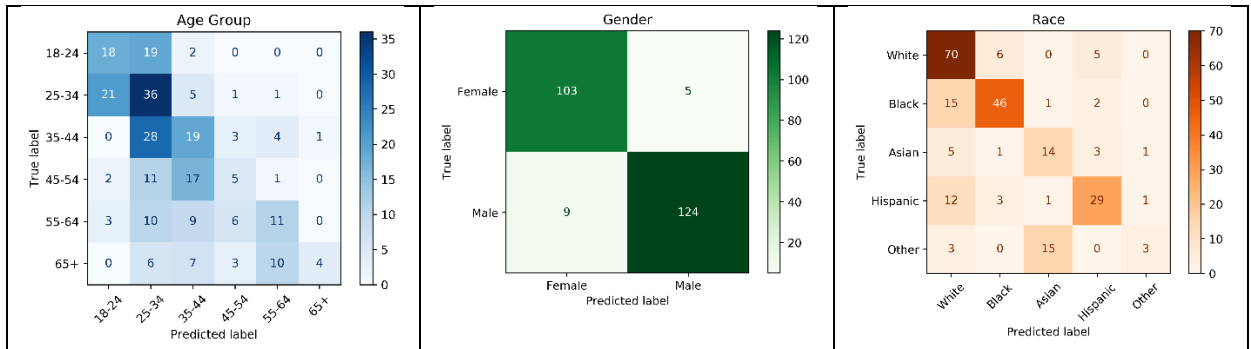
Figure 5. Confusion matrices for demographic inference under Condition C1 (names only)

## 4.2 Condition 2 (C2): Names + Short Description

In the second condition, users' short self-descriptions (biographical text) were added alongside usernames and display names. This introduces contextual linguistic cues—such as occupation, education, lifestyle, or personal interests—that may implicitly encode demographic information. Compared with C1, performance in C2 improves across all demographic dimensions (Table 3). Age accuracy increases from 0.35 to 0.44, one-off accuracy from 0.77 to 0.80, and MAE decreases from 10.01 to 8.57 years. The macro-F1 improves from 0.32 to 0.43, indicating more balanced age classification across groups. Gender performance remains stable (accuracy and macro-F1 = 0.94), suggesting that short descriptions do not substantially alter gender inference, likely because name-based cues already dominate this signal. Race/ethnicity shows modest improvement (accuracy = 0.72; macro-F1 = 0.62), indicating that biographical text adds useful ethnolinguistic and cultural context.

The confusion matrices (Figure 6) reveal noticeable improvement in age separability. Younger users (18–24, 25–34) and older users (55–64, 65+) are more clearly distinguished, reducing misclassification among adjacent bins. This likely results from age-indicative phrases in bios (e.g., "college student," "mother of three," "retired engineer"), which supplement the weak age signal in names. Gender accuracy remains consistently high (~94%) with minimal changes in error distribution. Race inference shows the largest qualitative gain, with fewer cross-group confusions for Black and White users and modest gains for Asian and Hispanic users. Self-descriptions often include culturally specific words or affiliations (e.g., "Latina," "HBCU," "Filipino American") that provide valuable ethnocultural cues. However, errors persist for the "Other" category and among users adopting culturally neutral or English-only descriptions, underscoring the continued challenge of inferring race in linguistically homogenized contexts.

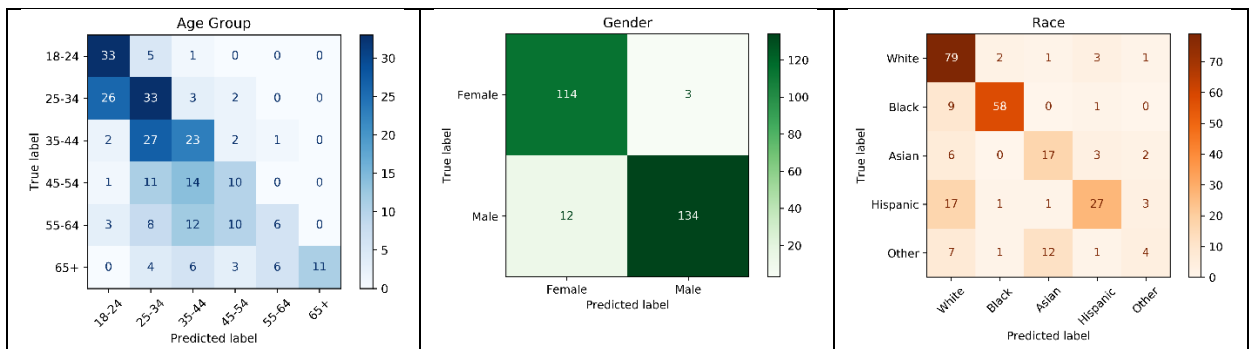
Figure 6. Confusion matrices for demographic inference under Condition C2 (names + short description)

## 4.3 Condition 3 (C3): Names + Description + One Tweet

In the third condition, a single tweet per user was added to the input, expanding the textual context. Tweets introduce spontaneous linguistic and behavioral patterns that capture users' communication style, vocabulary, and topic preferences, elements often correlated with demographic traits. Adding one tweet produces marked improvements across nearly all metrics (Table 3). Age prediction improves notably: accuracy rises from 0.44 to 0.51, one-off accuracy from 0.80 to 0.86, and MAE drops from 8.57 to 7.11 years. The macro-F1 increases from 0.43 to 0.49, suggesting stronger balance across age categories. Gender inference remains strong (accuracy and macro-F1 = 0.95), confirming the robustness of name-based cues while benefiting marginally from linguistic signals. Race/ethnicity achieves one of the largest relative gains, with accuracy improving from 0.72 to 0.77 and macro-F1 from 0.62 to 0.69. Joint accuracy rises substantially from 0.30 to 0.40, underscoring the integrative power of even a short behavioral text sample.

The confusion matrices (Figure 7) show further reductions in cross-age misclassifications, particularly between young and middle-aged groups. Tweets provide style markers (e.g., slang, emoji use, tone, topical focus) that correlate with generational cohorts, improving age sensitivity even with minimal text. Gender performance remains stable (≈95%), while race inference exhibits the most visible improvement—reduced confusion among Asian, Hispanic, and Other users, and higher within-group precision. Tweets contribute lexical and cultural markers such as ethnic references, language mixing (e.g., Spanglish, Tagalog), or identity expressions that enhance discrimination among underrepresented groups. Nevertheless, the "Other" category remains diffuse due to its inherent heterogeneity.

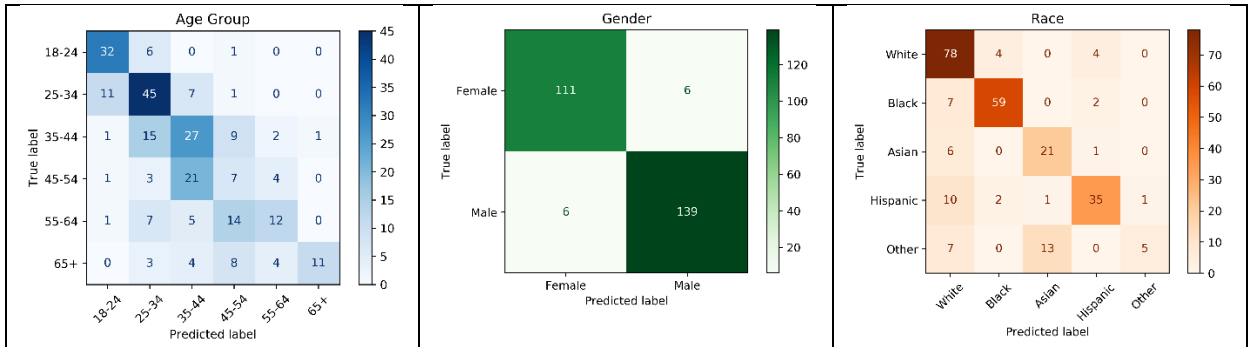

Figure 7. Confusion matrices for demographic inference under Condition C3 (names + description + one tweet).

## 4.4 Condition 4 (C4): Names + Description + Three Tweets

In the fourth condition, three tweets per user were included to test whether more textual input improves demographic inference beyond C3. As summarized in Table 3, adding two additional tweets yields incremental but modest gains. Age accuracy improves slightly from 0.51 to 0.52, MAE decreases from 7.11 to 6.92 years, and macro-F1 increases from 0.49 to 0.50. Gender accuracy rises slightly from 0.95 to 0.97, confirming stable and robust gender cues across linguistic samples. Race/ethnicity shows moderate improvement, with accuracy increasing from 0.77 to 0.80 and macro-F1 from 0.69 to 0.72. Joint accuracy improves marginally from 0.40 to 0.41, reflecting diminishing returns as additional textual information is added.

The confusion matrices indicate that age prediction patterns remain nearly identical to those in C3, suggesting that most age-related linguistic cues are captured from a single tweet. The slight reduction in

MAE likely reflects improved confidence or correction of a few outlier predictions rather than greater separability. Gender inference remains highly stable, while race classification continues to improve modestly, fewer errors for Black and Hispanic users and slightly better precision for Asian and Other groups. The additional tweets likely provide more consistent exposure to cultural and linguistic features that strengthen group-level distinctiveness, though with smaller marginal gains than the transition from C2 to C3.

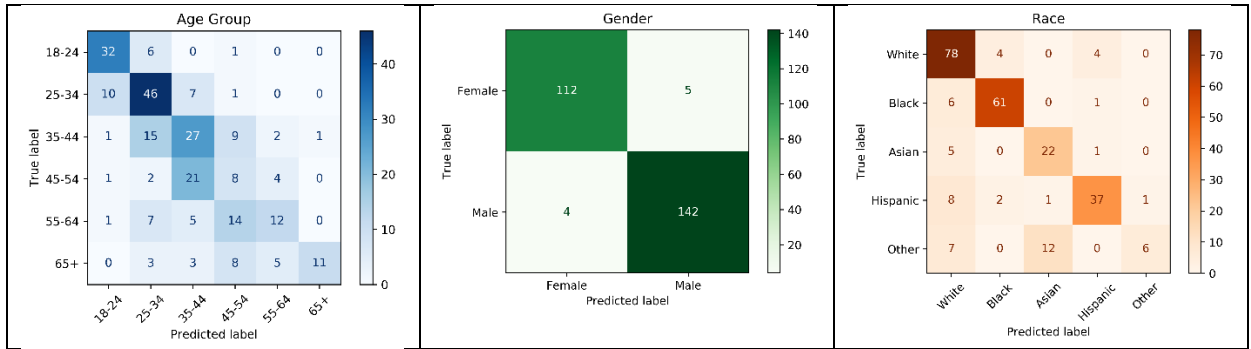

Figure 8. Confusion matrices for demographic inference under Condition C4 (names + description + three tweets)

## 4.5   Condition 5 (C5): Names + Description + Tweets + Image (Multimodal)

In the final condition, users' profile images were incorporated, introducing rich visual features such as facial attributes, apparent age, gender presentation, skin tone, and background context. The inclusion of visual data produces substantial performance gains (Table 3). Age accuracy rises from 0.52 to 0.58, one-off accuracy increases to 0.90, and MAE decreases from 6.92 to 5.54 years. The macro-F1 improves from 0.50 to 0.57, indicating more balanced classification across age groups. Gender accuracy reaches 0.98, the highest of all conditions, reflecting the strong visual cues for gender presentation. Race/ethnicity inference also improves significantly (accuracy = 0.85; macro-F1 = 0.79), confirming that visual information helps resolve the ambiguity of English-language names and text. Joint accuracy rises from 0.41 to 0.49, an 8% absolute gain, marking a substantial step toward holistic demographic prediction.

As shown in Figure 9, the age confusion matrix demonstrates improvement across nearly all categories: younger and older users become more distinguishable, and overall prediction dispersion narrows. The MAE reduction from 6.9 to 5.5 years indicates that visual cues allow the model to anchor predictions more precisely, correcting many adjacent-bin misclassifications seen in earlier stages. Gender inference reaches near-perfect accuracy (98%), as visual appearance provides clear and consistent cues that reinforce textual and nominal signals. Remaining errors likely correspond to non-personal images or cases where visual presentation diverges from conventional binary gender norms. Race prediction also shows qualitative improvement, with lower misclassification rates for Asian, Hispanic, and Other groups, and stronger diagonal dominance in the confusion matrix. Visual cues such as skin tone and facial structure supplement weak ethnolinguistic markers from text, though residual confusion within the "Other" category persists, reflecting the diversity and ambiguity of racial identity representation online.

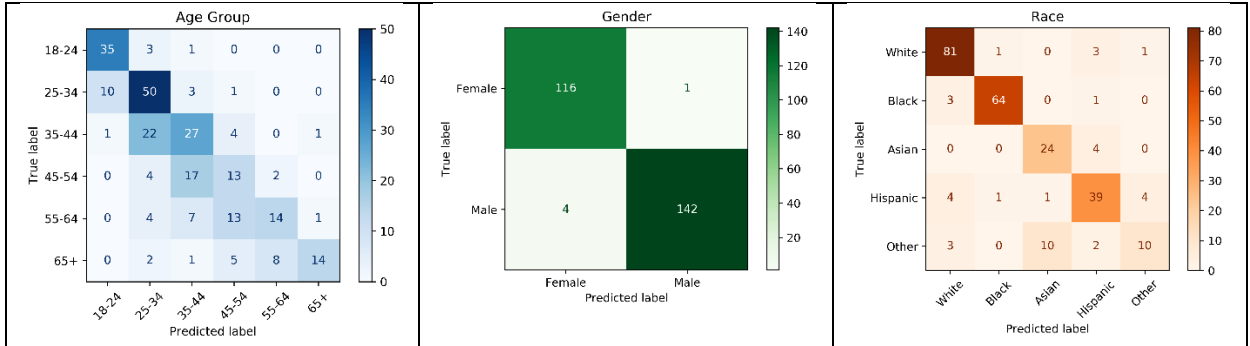

Figure 9. Confusion matrices for demographic inference under Condition C5 (names + description + tweets + profile image)

The final multimodal condition (C5) demonstrates that visual features substantially enhance demographic inference, particularly for age and race, where textual signals are inherently limited. By combining nominal, linguistic, and visual modalities, the model achieves its highest accuracy and lowest error across all demographic dimensions.

Table 3. Performance of GPT-5 across progressive input conditions (C1–C5) for demographic inference tasks.

| Cond. | Age Acc (%) | Age 1-off (%) | Age MAE (yrs) | Age Macro-F1 | Gender Acc (%) | Gender Macro-F1 | Race Acc (%) | Race Macro-F1 | Joint Acc (%) |
|---|---|---|---|---|---|---|---|---|---|
| C1 | 0.35 | 0.77 | 10.01 | 0.32 | 0.94 | 0.94 | 0.69 | 0.59 | 0.23 |
| C2 | 0.44 | 0.80 | 8.57 | 0.43 | 0.94 | 0.94 | 0.72 | 0.62 | 0.30 |
| C3 | 0.51 | 0.86 | 7.11 | 0.49 | 0.95 | 0.95 | 0.77 | 0.69 | 0.40 |
| C4 | 0.52 | 0.86 | 6.92 | 0.5 | 0.97 | 0.97 | 0.80 | 0.72 | 0.41 |
| C5 | 0.58 | 0.90 | 5.54 | 0.57 | 0.98 | 0.98 | 0.85 | 0.79 | 0.49 |

Figure 10 illustrates how each demographic inference metric evolves as progressively richer inputs are added. Overall, the figure shows a consistent upward trend across nearly all metrics, confirming the progressive enrichment effect of additional input modalities. Age-related metrics (accuracy, one-off accuracy, and macro-F1) exhibit the steepest improvements, rising sharply from C1 to C3, then stabilizing before a final increase in C5. This pattern suggests that linguistic and visual cues jointly enhance the model's ability to capture generational language and appearance signals. Gender metrics remain consistently high across all conditions, with only marginal gains as visual features are added, indicating that gender cues are strongly encoded in both names and facial features. Race metrics show a steady and continuous improvement, reflecting the combined contribution of name-based, linguistic, and visual information in capturing ethnocultural characteristics. Finally, joint accuracy increases almost linearly from C1 to C5, highlighting the additive value of each modality. The largest performance gains occur in the transitions from C2 to C3 and C4 to C5, demonstrating the substantial contributions of adding tweet content and visual data to the overall reliability of demographic inference. Together, these trends confirm the effectiveness of the progressive multimodal strategy in improving both the robustness and fairness of demographic inference.

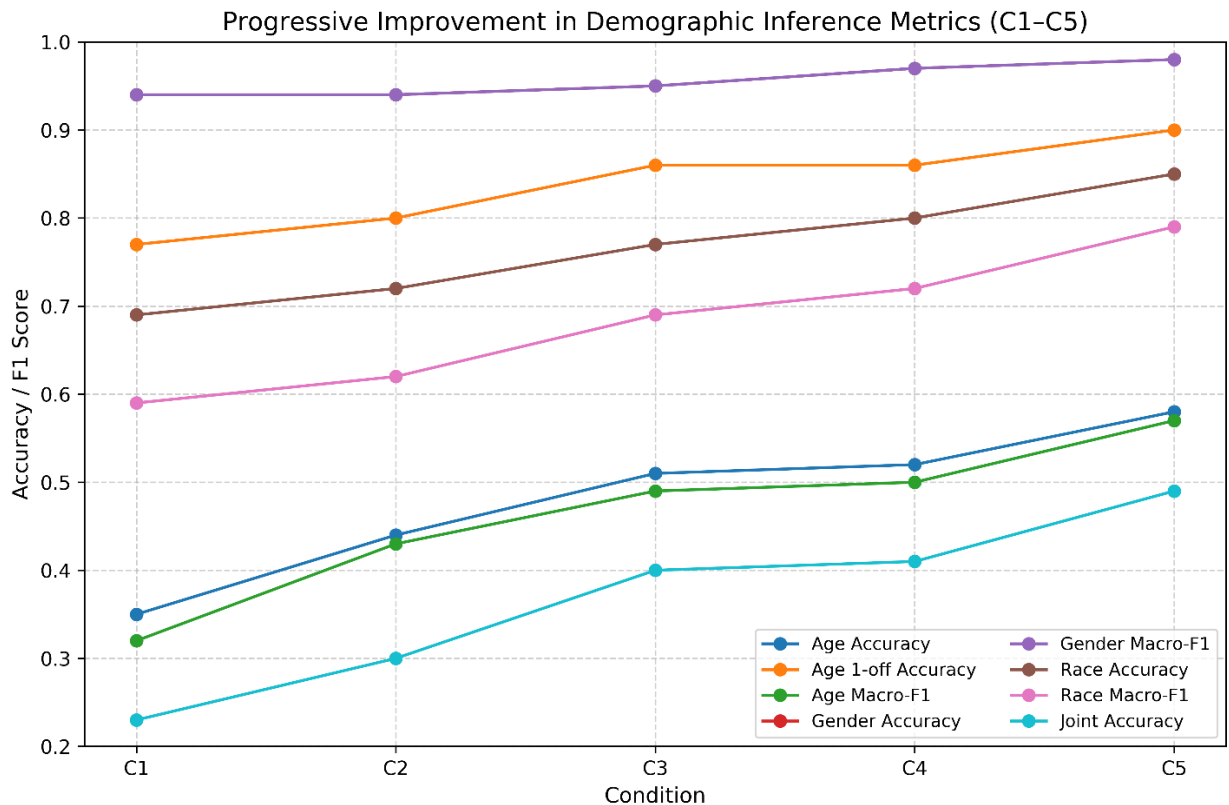

Figure 10. Performance trends of GPT-5 across progressive input conditions (C1–C5) for demographic inference

To further evaluate the performance of the proposed approach, we compared GPT-5 with several existing demographic inference models, including M3(Wang et al., 2019), GenderPerformer(Wang & Jurgens, 2018), and Jaech & Ostendorf(Jaech & Ostendorf, 2015). As noted earlier, there are currently no existing methods capable of jointly inferring age, gender, and race, which highlights the pioneering contribution of this study.

Among the comparison methods, M3 can infer both age and gender using information from usernames, display names, profile descriptions, and profile images. However, the M3 model defines age categories differently, specifically 19–29, 30–39, and ≥40, which are not directly comparable to our original six-group age classification. To ensure a fair comparison, we adjusted GPT-5's input configuration and output categories to match M3's settings, using only names, descriptions, and profile images as inputs and applying the same three age groups for output. For GenderPerformer and Jaech & Ostendorf, both methods predict gender based solely on names. Accordingly, GPT-5 was also evaluated under the same input condition (username only) for gender inference.

Table 4 summarizes the comparative results across all methods. GPT-5 consistently outperforms all existing methods under equivalent input conditions. When using only names, descriptions, and profile images, GPT-5 achieves higher accuracy than M3 for both age and gender inference. Notably, GPT-5's age accuracy in this setting (0.66) even exceeds that of the multimodal condition C5, primarily because the three-group classification reduces category overlap and error margins. When restricted to name-only inputs, GPT-5 again outperforms both GenderPerformer and Jaech & Ostendorf, achieving a gender accuracy of 0.91, compared with 0.84 and 0.78, respectively. This result demonstrates GPT-5's strong ability to capture gender distinctions from name morphology and linguistic patterns alone.

Overall, these findings emphasize that multimodal foundation models like GPT-5 can achieve superior accuracy with minimal input and no task-specific training, highlighting their potential as a general-purpose framework for demographic inference. This performance advantage reflects both the model's broad contextual understanding and its capacity to integrate heterogeneous information sources efficiently.

Table 4. Comparison of GPT-5 with existing demographic inference models under matched input settings.

| Model | Input | Age Acc | Age Macro-F1 | Gender Acc | Gender Macro-F1 | Race Acc | Race Macro-F1 | Joint Acc |
|---|---|---|---|---|---|---|---|---|
| GPT-5 | Username, Display Name + Short Description + Profile Image | 0.66 | 0.65 | 0.98 | 0.98 | 0.84 | 0.77 | 0.57 |
| M3 | Username, Display Name + Short Description + Profile Image | 0.51 | 0.49 | 0.94 | 0.94 | — | — | 0.40 |
| GPT-5 | Username Only | — | — | 0.91 | 0.91 | — | — | — |
| GenderPerformer | Username Only | — | — | 0.84 | 0.84 | — | — | — |
| Jaech & Ostendorf | Username Only | — | — | 0.78 | 0.78 | — | — | — |

## 5  Conclusion

This study systematically evaluated the capability of GPT-5, a state-of-the-art multimodal foundation model, to infer demographic attributes, including age, gender, and race, from social media profiles on X. Using a progressively enriched input design, we examined how different modalities contribute to the model's inference performance.

Results show a consistent and monotonic improvement across all conditions, confirming the additive value of multimodal information. Name-based inputs alone provided strong gender cues but weak age and race signals, reflecting the limitations of linguistic morphology for capturing demographic diversity. The inclusion of short bios and tweets introduced contextual and behavioral information, significantly improving accuracy for age and race inference. The largest performance gains occurred when tweet content and profile images were added, demonstrating the complementary strengths of linguistic and visual modalities. Visual cues, in particular, substantially reduced age and race ambiguity, leading to the highest overall accuracy and lowest error across all dimensions.

The study proves that multimodal foundation models can provide is promising solutions for demographic inference with social media data. The benchmarking confirms that GPT-5 delivers the best performance for demographic inference among all tested models. It achieves significantly higher accuracy than existing models but also offers greater flexibility in handling different input data types. This study also shared insights on effective prompting strategies to further enhance performance. Looking ahead, future work should extend this investigation to other social media platforms, such as Instagram and Reddit, where different modalities and communication styles may reveal new patterns in demographic expression. Future studies should also examine non-English and cross-cultural contexts to test the model's generalizability. In addition, research is needed to assess potential biases and ethical implications of large-scale demographic inference, ensuring that multimodal foundation models are applied responsibly in social data analytics and computational social science.